\documentclass{aa}

\usepackage{graphicx}
\usepackage{psfig}

\begin{document}

   \title{Galaxies in Southern Bright Star Fields\thanks{Based on 
          observations obtained at the European Southern 
          Observatory, Chile, for programmes 66.A-0361 and 
          68.A-0440.  Note: the entirety of Table \ref{t-src} is 
	  only available in electronic form at the CDS via anonymous 
	  ftp to cdsarc.u-strasbg.fr (130.79.128.5)
	  or via http://cdsweb.u-strasbg.fr/cgi-bin/qcat?J/A+A/.}}

   \subtitle{I. Near-infrared imaging}

   \author{Andrew J. Baker\inst{1}
	\and
	Richard I. Davies\inst{1}
	\and
	Matthew D. Lehnert\inst{1}
	\and
	Niranjan A. Thatte\inst{1,2}
	\and
	William D. Vacca\inst{1}
	\and
	Olivier R. Hainaut\inst{3}
	\and
	Matt J. Jarvis\inst{2,4}
	\and
	George K. Miley\inst{4}
	\and
	Huub J.~A. R{\" o}ttgering\inst{4}
	}

   \offprints{A.~J. Baker, {\tt ajb@mpe.mpg.de}}

   \institute{Max-Planck-Institut f{\" u}r extraterrestrische Physik \\
              Postfach 1312, D-85741 Garching, Germany\\
              \email{\{ajb,davies,mlehnert,thatte,vacca\}@mpe.mpg.de}
	      \and
	      University of Oxford Astrophysics \\
	      Keble Road, Oxford OX1 3RH, United Kingdom \\
	      \email{mjj@astro.ox.ac.uk}
	      \and
              European Southern Observatory \\
              Alonso de Cordova 3107, Casilla 19001, Vitacura, Santiago, 
              Chile \\
              \email{ohainaut@eso.org}
              \and
	      Sterrewacht Leiden \\
	      Postbus 9513, 2300 RA Leiden, the Netherlands \\
	      \email{\{miley,rottgeri\}@strw.leidenuniv.nl}
	      }

   \date{Received 7 March 2003 / Accepted 2 June 2003}

   \abstract{
As a prerequisite for cosmological studies using adaptive optics techniques, 
we have begun to identify and characterize faint sources in the vicinity of 
bright stars at high Galactic latitudes.  The initial phase of this work has 
been a program of $K_\mathrm{s}$ imaging conducted with SOFI at the ESO NTT.  
From observations of 42 southern fields evenly divided between the spring and 
autumn skies, we have identified 391 additional stars and 1589 galaxies lying 
at separations $\Delta \theta \leq 60\arcsec$ from candidate guide stars 
in the magnitude range $9.0 \leq R \leq 12.4$.  When analyzed as 
a ``discrete deep field'' with $131\,\mathrm{arcmin}^2$ area, our dataset 
gives galaxy number counts that agree with those derived previously over the 
range $16 \leq K_\mathrm{s} < 20.5$.  This consistency indicates that in the 
aggregate, our fields should be suitable for future statistical studies.  We 
provide our source catalogue as a resource for users of large telescopes in 
the southern hemisphere.

}

   \maketitle

\section{Introduction}

Understanding the mechanisms that drive galaxy formation and evolution is a 
central objective of modern astrophysics.  While the structures and dynamics 
of nearby galaxies constitute a valuable fossil record, these are much less 
informative than the properties of galaxies observed at the epochs of their 
formation and early evolution.  By studying galaxies' structures, dynamics, 
fundamental scaling laws, and distributions and rates of star formation over a 
range in redshift, it is possible to trace the evolving role and subtle 
interactions of key physical processes such as feedback, mass loss, merging, 
and secular evolution, among others.  Disentangling these processes in nearby 
systems is already extremely difficult.  Attempts to extend this effort to 
newly-formed galaxies require data whose high spatial resolution and dynamic 
range can only be achieved by 8m-class telescopes using adaptive optics (AO) 
techniques to work at or near their diffraction limits.

AO relies on the detection of, and compensation for, the distortions of 
wavefronts from a bright pointlike object.  For science targets that are 
themselves too faint to permit wavefront sensing, or too diffuse to sustain 
efficient wavefront correction (see Rousset \cite{rous94}), a bright ($V \leq 
13$), nearby ($\Delta \theta \leq 30\arcsec$) natural guide star (NGS) is 
required.  Exactly how bright and nearby depends on the local atmospheric 
conditions; the above values correspond to the typical regime in which the 
wavefront sensor is not photon-limited and the off-axis Strehl ratio at 
$2.2\,\mathrm{\mu m}$ is not too severely reduced (e.g., Le~Louarn et al. 
\cite{lelo98}).  In the absence of a nearby NGS, higher-order wavefront 
corrections using a laser guide star (LGS) can suffice, provided a moderately 
bright ($V \sim 16$) reference star lies relatively close ($\Delta \theta \leq 
90\arcsec$) to the science target for determination of the lowest-order 
tip-tilt correction.  Switching from LGS to NGS mode trades a gain in sky 
coverage for a loss in the highest possible Strehl ratio, however, since the 
LGS samples atmospheric turbulence in a conical (rather than cylindrical) 
volume.  The phase error due to such focus anisoplanatism increases with 
mirror diameter as $D^{5/6}$ (e.g., Le~Louarn et al. \cite{lelo98}, and 
references therein).  Until the advent of multiple-LGS systems that can 
correct for this effect (e.g., Tallon \& Foy \cite{tall90}), AO systems on 
8m-class telescopes will therefore reach the highest Strehl ratios only when 
operating in NGS mode.

In order to conduct cosmological studies with the present generation of AO 
technology (e.g., Larkin et al. \cite{lark00}; Davies et al. \cite{davi01}; 
Glassman et al. \cite{glas02}; Davies et al. \cite{davi03}), it is necessary 
to identify distant galaxies in the vicinity of bright guide stars.  
Unfortunately, most existing surveys either avoid bright stars (e.g., the 
Hubble Deep Field), or like DENIS and 2MASS (Epchtein et al. \cite{epch97}; 
Skrutskie et al. \cite{skru97}) are too shallow to detect very many 
targets at cosmological distances.  We have therefore begun a program to 
characterize the extragalactic sources lying close to bright stars at high 
Galactic latitudes.  Similar work underway by other authors (Larkin \& 
Glassman \cite{lark99}) has focused on fields easily observable from the Keck 
telescopes; our targets are specifically chosen to be at southern declinations 
suitable for observations with the NAOS-CONICA (Rousset et al. \cite{rous98}; 
Lenzen et al. \cite{lenz98}) and SINFONI (Thatte et al. \cite{that98}) 
instruments on the ESO Very Large Telescope (VLT).  

The initial phase of our program has been a campaign of (non-AO-assisted) 
near-IR imaging, since information about source magnitudes and surface 
brightness profiles will be most useful at the wavelength of eventual AO 
operation.  For deep imaging studies in particular, the availability of 
seeing-limited data for comparison will allow the empirical determination of 
the surface brightness selection effects influencing diffraction-limited data 
(e.g., whether a given faint ``point'' source is only the nucleus of an 
unremarkable extended galaxy).  Blind diffraction-limited imaging of random 
fields, in contrast, remains vulnerable to unknown biases and uncertainties in 
the surface densities of true point sources.  We have already followed up our 
near-IR imaging with optical imaging and multi-object spectroscopy of many 
of our fields; discussion of these data is deferred to future papers.

\section{Observations and data reduction}

\begin{table*}
 \caption[]{Bright star fields.  Columns are (1) field identifier by 
which the sources in Table \ref{t-src} are indexed (2) USNO-B1.0 
designation for the bright star (Monet et al. \cite{mone03}) (3) \& (4) 
USNO-B1.0 J2000.0 coordinates of the bright star, accurate to $\pm 
0.2\arcsec$ (5) \& (6) USNO-B1.0 proper motions of the bright star in 
$\mathrm{mas\,yr^{-1}}$, relative to the YS4.0 reference frame described by 
Monet et al. (\cite{mone03}) (7) \& (8) USNO-B1.0 $B$ and $R$ magnitudes 
of the bright star, averaged over two epochs and accurate to $\pm 0.3\,\mathrm{
mag}$ (9) $K_\mathrm{s}$ magnitude of the bright star, from our SOFI imaging 
(10) date(s) on which the field was observed with SOFI, as indexed in Table
\ref{t-obs} (11) total SOFI integration time, in minutes (12) note as to 
whether the field falls inside an EIS Wide patch or near an NVSS source.}
 \label{t-fields}
 \begin{tabular}{ccccrrrrrccc}
  \hline
  \noalign{\smallskip}
Field & USNO-B1.0 & $\alpha$\,(J2000.0) & $\delta$\,(J2000.0) & 
$\mu\,(\alpha)$ & $\mu\,(\delta)$ & $B$ & $R$ & $K_\mathrm{s}$ & Night(s) & 
$\Delta t$ & Note \\
(1) & (2) & (3) & (4) & (5) & (6) & (7) & (8) & (9) & (10) & (11) & (12) \\
  \noalign{\smallskip}
  \hline
  \noalign{\smallskip}
SBSF 01 & 0703-0002281 & 00 11 38.84 & $-$19 37 11.0 &  $+$8 &  $-$6 & 12.1 & 10.0 &  7.80 & 6     & 30 & NVSS  \\
SBSF 02 & 0601-0007980 & 00 44 31.88 & $-$29 52 30.3 & $+$24 &  $-$8 & 11.9 & 11.1 & 10.09 & 6     & 30 & EIS B \\
SBSF 03 & 0600-0008265 & 00 45 20.62 & $-$29 56 46.0 & $+$32 &  $+$6 & 12.9 & 12.1 & 11.11 & 6     & 30 & EIS B \\
SBSF 04 & 0604-0008038 & 00 45 28.05 & $-$29 31 40.1 & $-$12 &  $+$4 & 13.2 & 12.4 & 11.33 & 4 5   & 20 & EIS B \\
SBSF 05 & 0604-0008436 & 00 48 11.23 & $-$29 34 14.2 & $+$92 & $-$60 & 12.7 & 11.6 & 10.11 & 6     & 20 & EIS B \\
SBSF 06 & 0605-0009459 & 00 50 34.70 & $-$29 26 32.0 &     0 &     0 & 12.6 & 11.0 &  9.90 & 4 5   & 30 & EIS B \\
SBSF 07 & 0604-0008993 & 00 52 02.66 & $-$29 30 48.1 & $+$24 & $+$40 & 13.0 & 11.7 & 10.90 & 4 5   & 30 & EIS B \\
SBSF 08 & 0605-0009658 & 00 52 18.88 & $-$29 27 17.8 & $-$12 & $-$12 & 13.1 & 11.2 & 10.81 & 4 5   & 30 & EIS B \\
SBSF 09 & 0606-0009140 & 00 53 17.20 & $-$29 22 29.0 &  $-$2 & $-$26 & 12.2 & 11.4 & 10.21 & 6     & 30 & EIS B \\
SBSF 10 & 0605-0009783 & 00 53 17.95 & $-$29 25 22.6 & $+$26 & $-$42 & 12.0 & 11.1 &  9.97 & 5     & 30 & EIS B \\
SBSF 11 & 0656-0061729 & 05 36 33.64 & $-$24 19 56.5 &  $+$4 & $-$10 & 11.8 & 11.1 & 10.08 & 4 5   & 30 & EIS C \\
SBSF 12 & 0653-0066004 & 05 40 37.80 & $-$24 36 33.8 &  $+$2 & $-$20 & 12.1 & 11.5 & 10.26 & 4 5 6 & 30 & EIS C \\
SBSF 13 & 0654-0065981 & 05 41 49.93 & $-$24 35 32.2 &     0 &     0 & 10.9 & 10.8 & 10.46 & 5 6   & 20 & EIS C \\
SBSF 14 & 0767-0069418 & 06 07 06.34 & $-$13 13 37.1 & $+$12 & $-$26 &  9.8 &  9.0 &  7.99 & 5 6   & 20 & NVSS  \\
SBSF 15 & 0734-0203992 & 08 44 00.22 & $-$16 34 01.1 &  $-$4 &     0 & 11.3 & 11.8 & 11.00 & 1     & 30 &       \\
SBSF 16 & 0705-0208711 & 09 14 52.77 & $-$19 26 17.0 & $-$12 &     0 & 11.0 & 11.2 & 10.89 & 1     & 30 &       \\
SBSF 17 & 0683-0253521 & 09 47 44.79 & $-$21 37 12.7 &  $-$8 & $-$12 & 12.0 & 11.6 & 10.70 & 1     & 30 & EIS D \\
SBSF 18 & 0682-0261875 & 09 49 46.99 & $-$21 45 13.3 &     0 &     0 & 12.2 & 11.4 & 10.69 & 1     & 30 & EIS D \\
SBSF 19 & 0682-0262596 & 09 51 44.77 & $-$21 43 28.0 & $-$26 &  $+$2 & 13.2 & 11.9 & 11.06 & 2     & 30 & EIS D \\
SBSF 20 & 0696-0235495 & 09 51 50.64 & $-$20 20 14.8 &  $+$8 & $-$34 & 12.4 & 12.4 & 10.46 & 2     & 30 & EIS D \\
SBSF 21 & 0686-0238319 & 09 52 32.44 & $-$21 21 30.1 & $-$12 &  $+$2 & 12.7 & 11.9 & 11.33 & 2     & 30 & EIS D \\
SBSF 22 & 0696-0236655 & 09 55 14.78 & $-$20 20 03.3 & $-$12 &     0 & 12.4 & 10.6 & 10.23 & 1     & 30 & EIS D \\
SBSF 23 & 0528-0285326 & 10 09 22.50 & $-$37 11 54.8 &  $-$8 &     0 & 11.1 & 11.3 & 11.27 & 2     & 30 &       \\
SBSF 24 & 0599-0250386 & 10 40 26.20 & $-$30 00 36.5 & $-$14 &  $+$8 & 11.2 & 11.6 & 11.40 & 2     & 30 &       \\
SBSF 25 & 0577-0367919 & 12 24 02.32 & $-$32 14 35.7 & $-$18 &  $-$6 & 10.3 & 10.4 & 10.28 & 2 3   & 30 &       \\
SBSF 26 & 0593-0288091 & 12 49 55.31 & $-$30 38 18.7 & $-$26 &  $-$2 & 10.4 & 10.4 & 10.00 & 2 3   & 30 &       \\
SBSF 27 & 0582-0321519 & 12 55 37.48 & $-$31 42 41.3 & $-$10 & $-$24 & 11.4 & 11.6 & 10.97 & 1     & 30 &       \\
SBSF 28 & 0478-0346964 & 12 56 14.14 & $-$42 09 10.9 &  $-$8 &  $+$4 & 10.4 & 10.5 & 10.79 & 2 3   & 30 &       \\
SBSF 29 & 0580-0344020 & 13 03 18.33 & $-$31 54 27.3 &  $-$4 & $-$14 & 10.8 & 10.9 & 10.90 & 2     & 30 &       \\
SBSF 30 & 0550-0285689 & 13 04 37.76 & $-$34 56 30.8 &  $-$6 &  $+$6 & 11.4 & 11.7 & 10.80 & 1     & 30 &       \\
SBSF 31 & 0460-0297795 & 13 28 32.45 & $-$43 58 03.4 & $-$14 &  $-$2 & 11.7 & 12.1 & 10.57 & 1     & 30 &       \\
SBSF 32 & 0562-0303721 & 13 28 59.77 & $-$33 42 25.0 & $-$28 & $+$16 & 10.3 & 10.3 & 10.05 & 2     & 30 &       \\
SBSF 33 & 0568-0379911 & 13 40 31.54 & $-$33 07 52.6 & $-$12 & $-$16 & 10.7 & 10.8 & 10.76 & 2     & 30 &       \\
SBSF 34 & 0582-0339867 & 13 46 25.24 & $-$31 45 47.5 &  $+$8 &  $-$8 & 11.4 & 11.6 & 10.97 & 1     & 30 &       \\
SBSF 35 & 0745-0822065 & 21 00 25.58 & $-$15 26 12.4 & $-$14 & $-$12 & 12.1 & 10.4 &  8.40 & 5 6   & 30 & NVSS  \\
SBSF 36 & 0615-0931505 & 22 14 36.74 & $-$28 25 31.6 & $-$16 & $-$16 & 11.7 & 10.6 &  9.36 & 5     & 30 & NVSS  \\
SBSF 37 & 0501-0829581 & 22 43 04.43 & $-$39 49 29.3 & $+$28 & $-$22 & 12.8 & 11.8 & 10.55 & 6     & 30 & EIS A \\
SBSF 38 & 0498-0817047 & 22 47 06.77 & $-$40 10 01.3 & $+$18 &  $-$8 & 11.5 & 10.8 &  9.29 & 6     & 30 & EIS A \\
SBSF 39 & 0504-0833030 & 22 49 34.23 & $-$39 33 05.3 & $+$18 & $-$36 & 12.9 & 11.1 &  9.78 & 5     & 20 & EIS A \\
SBSF 40 & 0501-0831035 & 22 49 49.32 & $-$39 53 15.0 & $-$36 & $-$28 & 12.4 & 11.8 & 10.64 & 5     & 30 & EIS A \\
SBSF 41 & 0498-0817689 & 22 50 21.28 & $-$40 07 38.6 & $+$18 &     0 & 12.5 & 11.2 & 10.72 & 5     & 30 & EIS A \\
SBSF 42 & 0714-0852904 & 23 29 55.77 & $-$18 35 54.1 &  $+$6 &  $+$4 & 11.0 &  9.6 &  7.90 & 6     & 30 & NVSS  \\
  \noalign{\smallskip}
  \hline
  \noalign{\smallskip}
 \end{tabular}
\end{table*}

\begin{figure*}
\centerline{\psfig{file=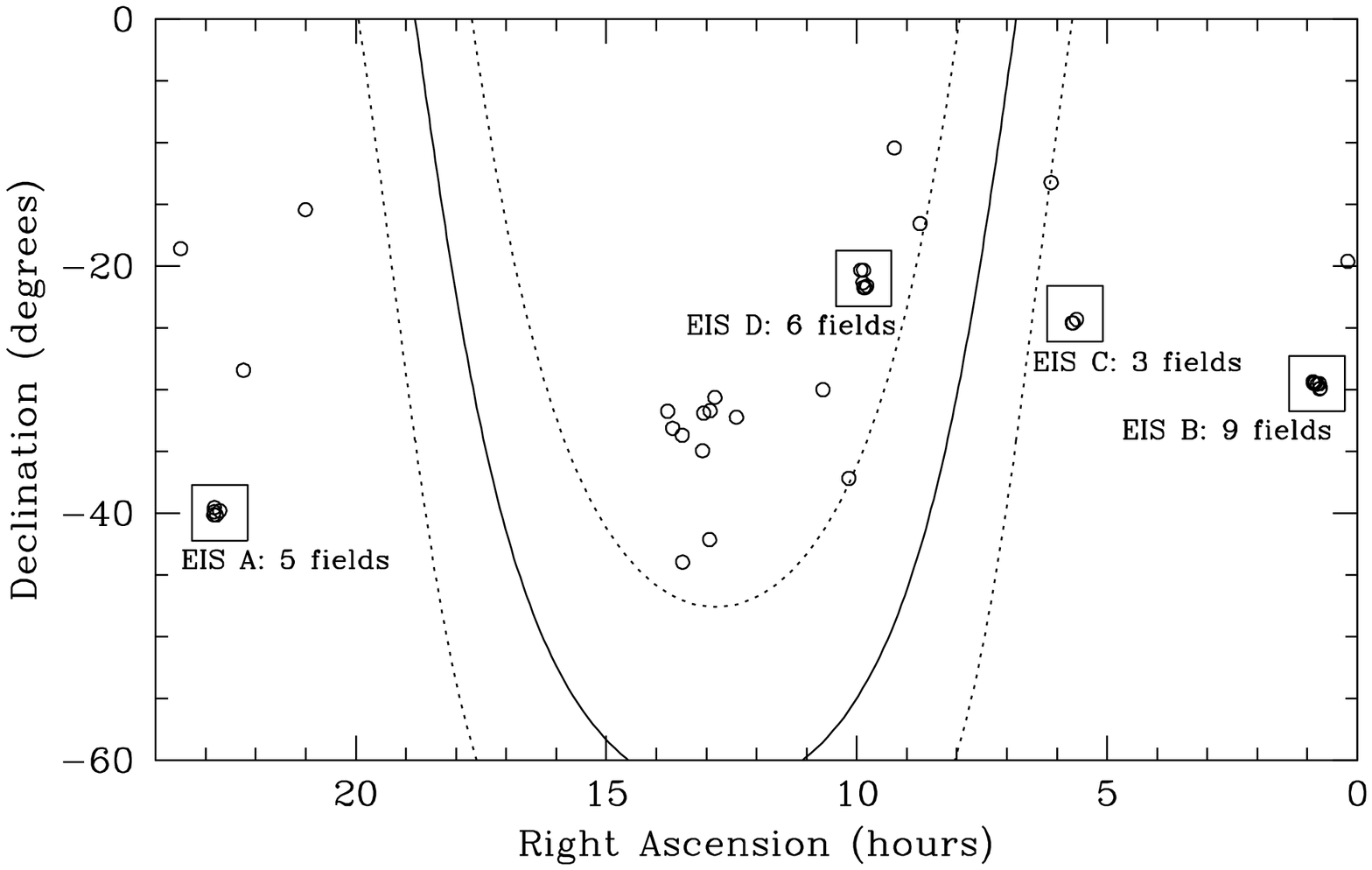,width=15cm}}
\caption{Positions of the 42 southern bright star fields observed in 
this survey, denoted by open circles.  The solid line shows the 
Galactic plane; the dotted lines to either side indicate Galactic 
latitude $|b|=15\degr$.  The four EIS Wide patches from which some 
of the fields were selected are indicated by boxes (not drawn to 
scale).}
\label{f-sky}
\end{figure*}

\subsection {Target selection}

Table \ref{t-fields} lists the identifications, coordinates, and magnitudes of 
the 42 bright stars that define our near-IR imaging sample; Fig. \ref{f-sky}
shows their distribution on the sky.  The majority of these targets were 
selected from the USNO-A2.0 catalogue (Monet et al. \cite{mone98}) 
according to criteria designed to ensure optimal performance of a near-IR 
science instrument assisted by a visible-wavelength (e.g., $0.45-1.0\,\mathrm{
\mu m}$ for NAOS-CONICA) wavefront sensor:
\begin{enumerate}

\item magnitude $10.3 \leq R \leq 12.4$, i.e., bright enough in the 
visible for wavefront sensing to be robust;

\item color $B-R \leq 1.1$, i.e., bluer than that of a type G1 star, in 
order to maximize the number of photons on the wavefront sensor while 
minimizing the amount of near-IR light scattered onto the science detector; 

\item Galactic latitude $|b| \geq 15\degr$, in order to minimize extinction 
and contamination by large numbers of foreground stars; and 

\item declination in the range $-44\degr \leq \delta \leq -13\degr$, so 
that AO observations from Cerro Paranal (at latitude $\sim -24\degr$) 
can be obtained at low airmass.

\end{enumerate}
Because the USNO-A2.0 catalogue's star-galaxy separation is not 
perfect (e.g., USNO-A2.0 0450-40377908 = \object{ESO\,436-G001}), our 
search was not entirely automated.  To check that favored objects were in fact 
stars, we used either the Tycho-2 catalog (H{\o}g et al. \cite{hog00}) or the 
publically available ESO Imaging Survey (EIS) data.  Visual inspection of the 
$I$-band images for EIS Wide patches A (Nonino et al. \cite{noni99}), B 
(Prandoni et al. \cite{pran99}), and C and D (Benoist et al. \cite{beno99}) 
allowed us to confirm that no additional bright stars or large foreground 
galaxies would block our view of more interesting background objects; we 
also tended to favor fields with one or more discernable $I$-band sources lying
within $30\arcsec$ of the central stars.  In a small number of cases, we 
relaxed our magnitude and color criteria to include brighter and redder stars 
located within $20\arcsec$ of radio sources detected in the NRAO VLA Sky 
Survey (NVSS: Condon et al. \cite{cond98}).  Our goal in imaging these fields 
(noted in the last column of Table \ref{t-fields}) was to take advantage 
of the fact that radio galaxies often trace overdensities in the large-scale 
structure, seemingly at all redshifts (e.g., Hill \& Lilly \cite{hill91}; Best 
\cite{best00}; Kurk et al. \cite{kurk00}; Pentericci et al. \cite{pent00}; 
McLure \& Dunlop \cite{mccl01}).  Subsequent to our observations, the 
release of the USNO-B1.0 catalogue (Monet et al. \cite{mone03}) allowed us to 
verify that only one of the stars in our sample (SBSF\,05) has a proper 
motion as large as $50\,\mathrm{mas\,yr^{-1}}$.\footnote{The preliminary 
photometry in the USNO-B1.0 catalogue shifts a few additional stars to $B - R 
> 1.1$, as Table \ref{t-fields} indicates.}

\subsection{Observations} \label{ss-obs}

We obtained our observations using SOFI (Moorwood et al. \cite{moor98}) on the 
ESO New Technology Telescope (NTT), over the course of six nights in February, 
March, and October 2001 (see Table \ref{t-obs}).  Conditions were photometric 
for two of the three nights in each season, allowing us to obtain good 
photometry for all of our fields; $K_\mathrm{s}$ seeing ranged from 
$0.5\arcsec$ to $1.0\arcsec$ with a median $\sim 0.7\arcsec$.  Our choice of 
observing strategy was guided by our desire to optimize the detection of faint 
sources at small angular separations from the bright star in each field.  We 
used SOFI's small-field objective lens, giving $0.145\arcsec \times 
0.145\arcsec$ pixels and a $2.47\arcmin \times 2.47\arcmin$ field of view; 
larger pixels would only have hindered modelling of point spread function 
(PSF) wings close to the star, while a larger field of view would only have 
yielded additional sources too far from the star to be useful for future AO 
studies.  In order to minimize memory effects in the HgCdTe array, we used 
short detector integration times (2--4\,s, depending on the magnitude of the 
star) and spent a total of one minute on source before dithering the telescope 
to a new position.  After imaging the field at ten such positions-- each 
located randomly within a ``jitter box'' of dimensions $40\arcsec \times 
40\arcsec$-- we switched to a different bright star field for another 
ten-frame sequence.  We would return to a particular field again later in the 
same (or a different) night, in order to allow the diffraction spikes of the 
(Alt-Az mounted) NTT to rotate on the sky and thus expose any faint objects 
lying underneath.  All of our fields were observed for a total of at least 20 
minutes, and most were observed for a total of 30 minutes (see Table 
\ref{t-fields}).  During each of the four photometric nights, we interspersed 
multiple observations of six or seven faint standard stars (Persson et al. 
\cite{pers98}) in order to determine the flux scale. 

\subsection{Data reduction}

\begin{table}
 \caption[]{Log of observations.}
 \label{t-obs}
 \begin{tabular}{ccc}
  \hline
  \noalign{\smallskip}
Night & Date & Conditions \\
(1) & (2) & (3) \\
  \noalign{\smallskip}
  \hline
  \noalign{\smallskip}
1 & 2001 Feb 15 & photometric \\
2 & 2001 Feb 17 & photometric \\
3 & 2001 Mar 07 & thin cirrus \\
4 & 2001 Oct 06 & thin cirrus \\
5 & 2001 Oct 07 & photometric \\
6 & 2001 Oct 08 & photometric \\
  \noalign{\smallskip}
  \hline
  \noalign{\smallskip}
 \end{tabular}
\end{table}

We reduced the data using custom scripts within the NOAO IRAF package 
(Tody \cite{tody93}).  For each set of ten consecutive frames, we 
made an estimate of the (sky and dark current) background by masking out 
objects and-- after further rejection of high and low pixels according to an 
estimated variance-- taking a median.  Following background subtraction, each 
frame was divided by a dome flat field (the difference of lamp on and lamp off 
frames) that had been corrected for the dependence of the array shade pattern 
on incident flux.\footnote{This type of ``special'' dome flat is described in 
Version 1.3 of the SOFI User's Manual.}  Values for bad pixels were 
then interpolated.  We next addressed the problem of low-level array cross 
talk, which resulted from the bright star in the field and manifested 
as excess counts in two vertical stripes.  To make a more precise measurement 
of this rather weak effect, we coarsely aligned the ten frames, averaged them, 
and measured the median of each column in the mean image.  The resulting 
values were then subtracted from the corresponding columns in the individual 
unaligned frames.  Finally, we performed a fine alignment to compensate for 
irregular behavior by the image derotator during our runs.  Although the 
derotator always assumed the correct position for the initial frame in each 
set of ten, it often remained at the same position during the subsequent nine.
We therefore rotated and translated frames 2--10 in each set to align them 
with frame 1.  For the final image combination (of all 20--30 frames taken 
for each field), after rejection of high and low pixels according to an 
estimated variance, we calculated the mean of the remaining pixels at each 
point.  The angular resolution in the final images was calculated as the FWHM 
of a Moffat (\cite{moff69}) profile fit to the central star in each field.  

For each observing run separately, the conditions were sufficiently stable 
that we could use the same flux calibration scale for the entire dataset.
The zero point magnitudes were $K_\mathrm{s} = 22.27 \pm 0.02$ for the 
February/March run and $K_\mathrm{s} = 22.34 \pm 0.03$ for the October run, 
with uncertainties reflecting the scatter from multiple standard stars 
observed at a variety of times, airmasses, and positions on the detector 
during the course of each night.  These and all other magnitudes in this 
paper are Vega-relative.

After measuring the $K_\mathrm{s}$ magnitudes of the central stars (see Table 
\ref{t-fields}), we wanted-- to the extent possible-- to remove all traces 
of them from the images.  To eliminate the (approximately) azimuthally 
symmetric part of each star's PSF, we fit ellipses to its isophotes.  Inside 
the FWHM of the star's surface brightness profile, we allowed the position 
angle and ellipticity to be free parameters; outside the FWHM, these were 
fixed to the values at the FWHM to permit isophote-fitting to the PSF wings 
out to a radius of $\sim 15\arcsec$.  Deviating pixels (i.e., those that could 
in principle be due to faint underlying sources) were rejected in these fits.  
The azimuthally {\it asymmetric} diffraction spikes in the star's PSF had in 
general already been excised at the earlier stage of image combination, where 
their different orientations for the different sets of ten frames allowed 
them to be eliminated via high pixel rejection.  For the small minority of 
fields where this was not possible or that centered on very bright stars, no 
satisfactory method could be found to remove the diffraction spikes that did 
not also compromise the quality of the data.

\section{Analysis}

\begin{table*}
 \caption[]{Sources in bright star fields identified by SExtractor;  
columns are described in Sect. \ref{ss-sex}.  A full version is 
available in electronic form at the CDS.}
 \label{t-src}
 \begin{tabular}{ccccccccc}
  \hline
  \noalign{\smallskip}
Source & R.~A. & Dec. & $\Delta \theta$ & $K_\mathrm{s}$ & Major axis & Minor 
axis & P.~A. & Stellarity \\
identifier & (J2000) & (J2000) & (arcsec) & (mag) & (arcsec) & (arcsec) & 
(degrees) & index \\
(1) & (2) & (3) & (4) & (5) & (6) & (7) & (8) & (9) \\
  \noalign{\smallskip}
  \hline
  \noalign{\smallskip}
SBSF 01$+$077$+$024 & 00 11 39.38 & $-$19 37 08.6 &  8.1 & 14.04 &  1.63 &  1.11 &  81 & 0.09 \\
SBSF 01$+$047$-$093 & 00 11 39.17 & $-$19 37 20.3 & 10.5 & 15.35 &  1.63 &  0.65 &  50 & 0.03 \\
SBSF 01$-$132$-$076 & 00 11 37.90 & $-$19 37 18.6 & 15.2 & 20.12 &  0.14 &  0.11 &  28 & 0.42 \\
SBSF 01$+$130$+$149 & 00 11 39.76 & $-$19 36 56.1 & 19.8 & 18.60 &  0.35 &  0.29 &  71 & 0.96 \\
SBSF 01$-$217$-$099 & 00 11 37.30 & $-$19 37 20.9 & 23.9 & 17.83 &  0.49 &  0.42 & 131 & 0.03 \\
. & . & . & . & . & . & . & . & . \\
. & . & . & . & . & . & . & . & . \\
. & . & . & . & . & . & . & . & . \\
SBSF 42$-$057$-$480 & 23 29 55.36 & $-$18 36 42.0 & 48.4 & 19.34 &  0.30 &  0.22 &  60 & 0.55 \\
SBSF 42$-$487$+$056 & 23 29 52.34 & $-$18 35 48.4 & 49.0 & 18.26 &  0.47 &  0.37 & 147 & 0.19 \\
SBSF 42$+$516$-$130 & 23 29 59.40 & $-$18 36 07.0 & 53.2 & 14.12 &  0.61 &  0.56 &  91 & 0.98 \\
SBSF 42$+$335$-$466 & 23 29 58.12 & $-$18 36 40.6 & 57.5 & 19.71 &  0.23 &  0.21 & 170 & 0.51 \\
SBSF 42$-$016$+$590 & 23 29 55.65 & $-$18 34 55.0 & 59.0 & 20.82 &  0.12 &  0.10 &  47 & 0.38 \\
  \noalign{\smallskip}
  \hline
  \noalign{\smallskip}
 \end{tabular}
\end{table*}

\subsection{Object identification} \label{ss-sex}

We used SExtractor (Bertin \& Arnouts \cite{bert96}) to generate object 
catalogues from our final, PSF-subtracted images.  Table \ref{t-src} 
lists the sources, for each of which we provide the following information:
\begin{enumerate}

\item a unique identifier that combines a field name from Table \ref{t-fields} 
with the source's right ascension and declination offsets from the central 
star in tenths of arcseconds (SBSF\,01$+$047$-$093, for example, is the object 
$4.7\arcsec$ east and $9.3\arcsec$ south of the bright star in field SBSF\,01);

\item its right ascension, estimated from the right ascension offset and 
columns (3) through (5) of Table \ref{t-fields};

\item its declination, estimated from the declination offset and 
columns (4) and (6) of Table \ref{t-fields};

\item its angular separation from the bright star; 

\item its $K_\mathrm{s}$ magnitude, using the SExtractor BEST definition that 
appears in simulations to be satisfactorily robust against aperture effects 
(V{\" a}is{\" a}nen et al. \cite{vais00}; cf. Martini \cite{mart01});

\item its semi-major axis, defined as the second-order moment (i.e., the 
intensity-weighted RMS) parallel to its major axis;

\item its semi-minor axis, defined as the second-order moment perpendicular to 
its major axis;

\item the position angle of its major axis, in degrees east of north; and 

\item its SExtractor stellarity index, ranging from 0 (diffuse) to 1 
(pointlike).

\end{enumerate}
Within Table \ref{t-src}, the bright star fields are listed in order of right 
ascension; within each field, the individual sources are then listed in order 
of increasing angular separation from the central star, out to a maximum of 
$\Delta \theta = 60\arcsec$.  The full catalogue includes 1980 sources; this 
total does not include the 42 bright stars on which the fields themselves are 
centered.  

\begin{figure}
\centerline{\psfig{file=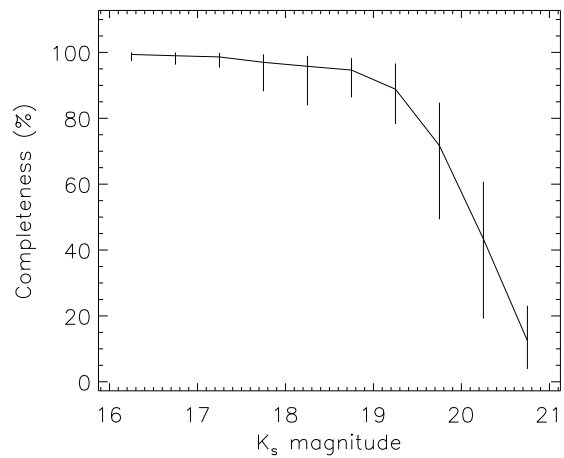,width=8.4cm}}
\caption{Completeness as a function of magnitude. The solid line is the mean 
over all 42 fields; the error bars indicate the dispersions between
fields, reflecting field-to-field variations in sensitivity.}
\label{f-cpl}
\end{figure}

\begin{figure}
\centerline{\psfig{file=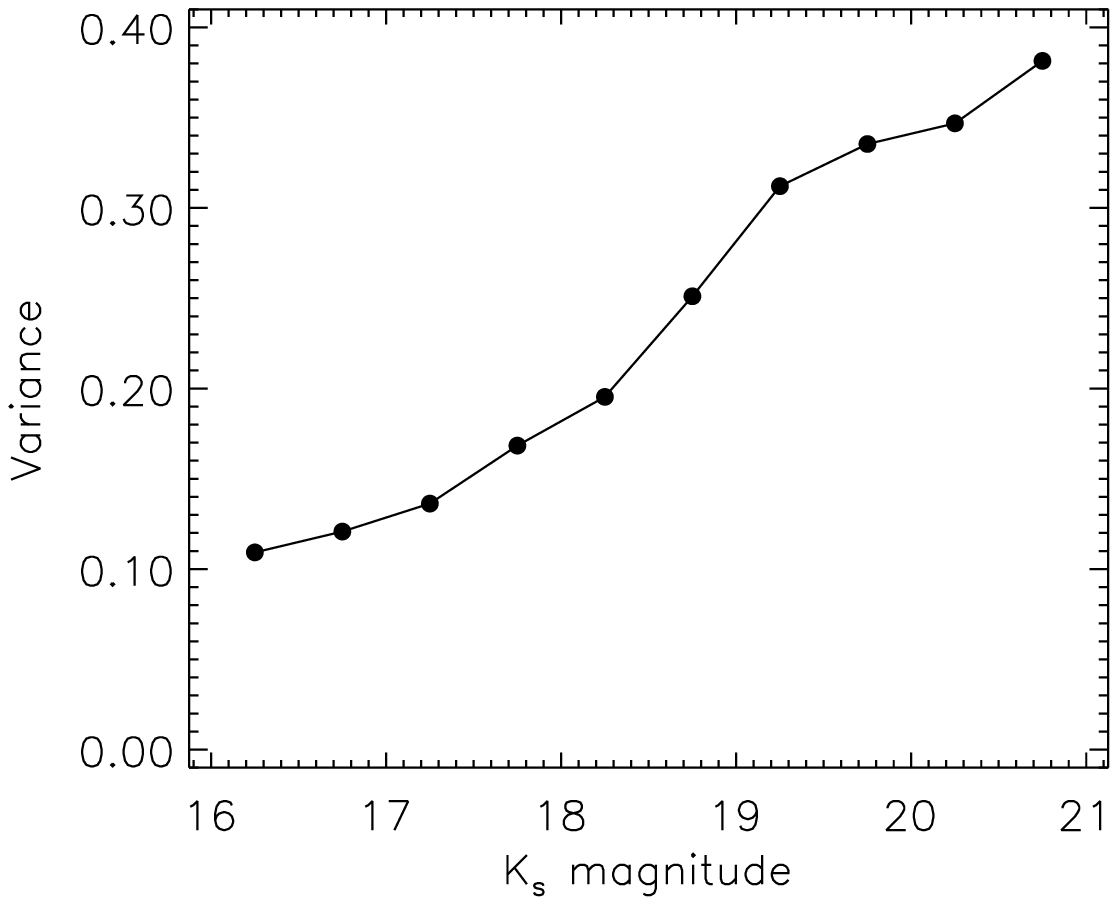,width=8.4cm}}
\caption{Uncertainty in our estimate of $K_\mathrm{s}$ magnitude as a function 
of true source magnitude.  As described in Sect. \ref{ss-comp}, we plot the 
variance of the distribution of measured magnitudes for each set of nominal 
magnitudes (i.e., of fictitious objects pasted into simulated images).}
\label{f-mag_sdv}
\end{figure}

We estimate that the offsets of each field's sources relative to the 
central star are accurate to within the size of a SOFI pixel, i.e., $\pm 
0.075\arcsec$.  Although this relative astrometry formally applies only to the 
2001.1 and 2001.8 epochs of our observations, the nominal proper motions 
of the reference stars are small enough that (with the exception of SBSF\,05) 
J2000.0 coordinates can essentially be computed from the field centers 
and source identifiers alone.  We note that the absolute coordinates 
listed in Table \ref{t-src} may be inaccurate if the central stars' true 
proper motions differ from the {\it relative} values listed in the USNO-B1.0 
catalogue and Table \ref{t-fields}, or if some of the (stellar) objects 
themselves have nonzero proper motions.

\subsection{Completeness and photometric accuracy} \label{ss-comp}

\begin{figure}
\centerline{\psfig{file=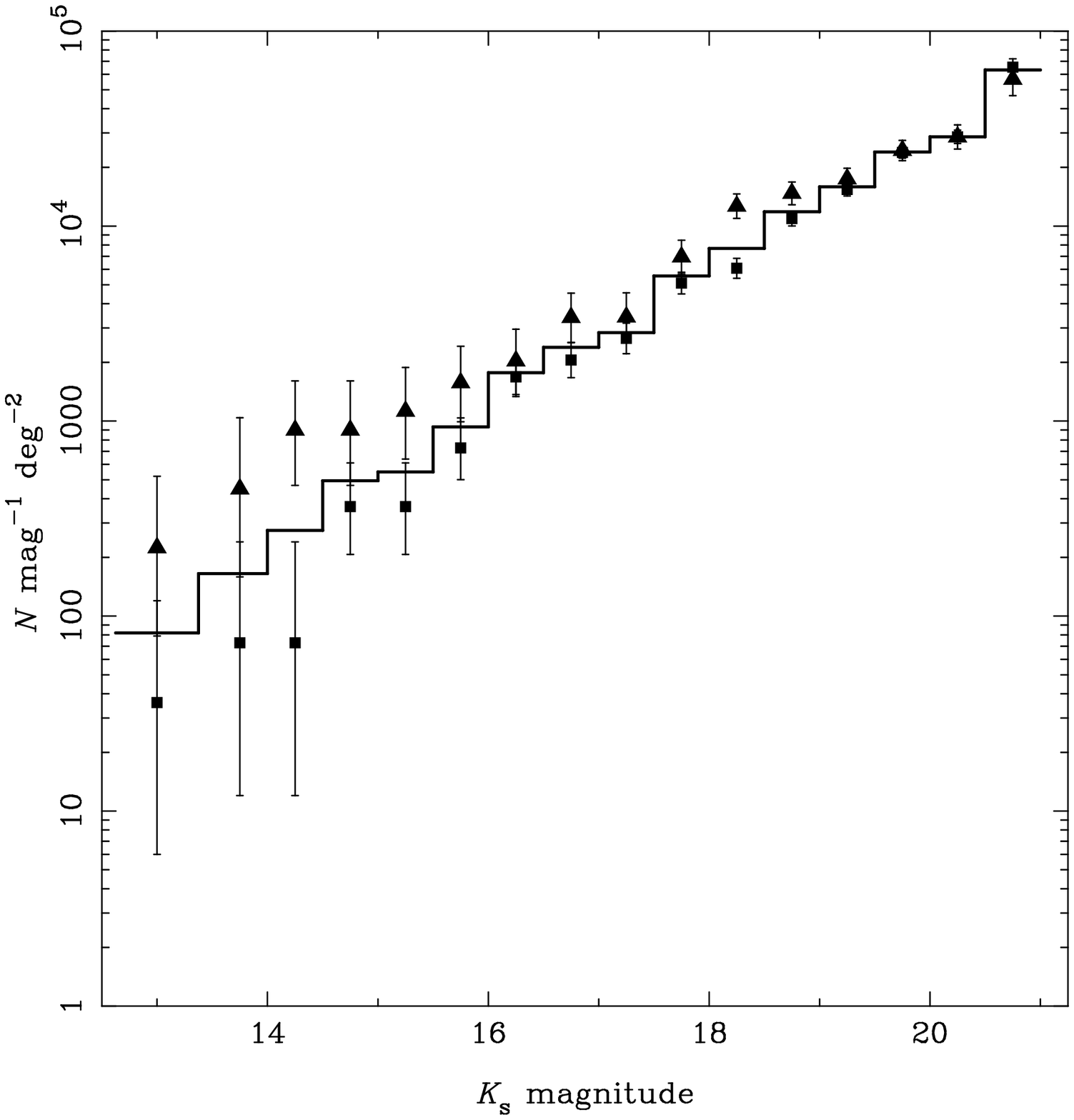,width=8.4cm}}
\caption{Differential galaxy number counts as a function of $K_\mathrm{s}$ 
magnitude.  Triangles and squares indicate completeness-corrected counts for 
the disjoint sets of sources with separations $5\arcsec \leq \Delta \theta 
< 30\arcsec$ and $30\arcsec \leq \Delta \theta \leq 60\arcsec$ from the 
bright stars (i.e., columns (6) and (7) in Table \ref{t-counts}), 
respectively.  The thick solid line shows the relation (also 
completeness-corrected) when these two subsets are combined (column (8) in 
Table \ref{t-counts}).}
\label{f-mags}
\end{figure}

\begin{table*}
 \caption[]{Galaxy number counts as a function of $K_\mathrm{s}$ magnitude.  
Column (1) indicates the magnitude bin.  Column (2) list the corresponding 
completeness $f_\mathrm{c}$ (when reliable).  Columns (3), (4), and (5) list 
the raw numbers of galaxies with source separations $\Delta \theta < 
5\arcsec$, $5\arcsec \leq \Delta \theta < 30\arcsec$, and $30\arcsec \leq 
\Delta \theta \leq 60\arcsec$, respectively.  Columns (6), (7), and (8) list 
the completeness-corrected number {\it counts} of galaxies per magnitude per 
square degree for source separations $5\arcsec \leq \Delta \theta < 
30\arcsec$, $30\arcsec \leq \Delta \theta \leq 60\arcsec$, and $5\arcsec \leq 
\Delta \theta \leq 60\arcsec$, respectively.  We do not list number counts 
for the two faintest magnitude bins due to the substantial uncertainty 
in the true $f_\mathrm{c}$ at these levels.  Uncertainties are calculated from 
Poissonian statistics following Gehrels (\cite{gehr86}).}
 \label{t-counts}
 \begin{tabular}{cccccccc}
  \hline
  \noalign{\smallskip}
$K_\mathrm{s}$ & $f_\mathrm{c}$ & $N_\mathrm{raw}\,(0,5)$ & 
$N_\mathrm{raw}\,(5,30)$ & $N_\mathrm{raw}\,(30,60)$ & 
$N\,(5,30)$ & $N\,(30,60)$ & $N\,(5,60)$ \\
(1) & (2) & (3) & (4) & (5) & (6) & (7) & (8) \\
\hline
12.5 -- 13.5 & 1.000 &  0 &   2 &   1 & $   224^{+297}_{-145}$  & $     36^{+84}_{-30}$   & $     82^{+80}_{-44}$   \\
13.5 -- 14.0 & 1.000 &  0 &   2 &   1 & $   449^{+592}_{-290}$  & $    73^{+167}_{-61}$   & $   165^{+160}_{-90}$   \\
14.0 -- 14.5 & 1.000 &  0 &   4 &   1 & $   898^{+710}_{-430}$  & $    73^{+167}_{-61}$   & $   275^{+185}_{-119}$  \\
14.5 -- 15.0 & 1.000 &  0 &   4 &   5 & $   898^{+710}_{-430}$  & $   364^{+246}_{-157}$  & $   494^{+226}_{-161}$  \\
15.0 -- 15.5 & 1.000 &  0 &   5 &   5 & $  1122^{+760}_{-484}$  & $   364^{+246}_{-157}$  & $   549^{+235}_{-170}$  \\
15.5 -- 16.0 & 1.000 &  0 &   7 &  10 & $  1571^{+847}_{-579}$  & $   728^{+310}_{-227}$  & $   934^{+286}_{-224}$  \\
16.0 -- 16.5 & 0.994 &  0 &   9 &  23 & $  2033^{+928}_{-665}$  & $  1684^{+429}_{-349}$  & $  1769^{+371}_{-311}$  \\
16.5 -- 17.0 & 0.990 &  1 &  15 &  28 & $ 3402^{+1124}_{-869}$  & $  2058^{+466}_{-387}$  & $  2387^{+422}_{-363}$  \\
17.0 -- 17.5 & 0.986 &  0 &  15 &  36 & $ 3415^{+1130}_{-872}$  & $  2656^{+521}_{-440}$  & $  2842^{+457}_{-397}$  \\
17.5 -- 18.0 & 0.970 &  0 &  30 &  68 & $ 6943^{+1514}_{-1261}$ & $  5100^{+697}_{-616}$  & $  5552^{+619}_{-560}$  \\
18.0 -- 18.5 & 0.958 &  1 &  54 &  80 & $12655^{+1968}_{-1716}$ & $  6076^{+758}_{-678}$  & $  7686^{+723}_{-663}$  \\
18.5 -- 19.0 & 0.946 &  0 &  62 & 142 & $14714^{+2117}_{-1863}$ & $ 10921^{+996}_{-915}$  & $ 11849^{+890}_{-829}$  \\
19.0 -- 19.5 & 0.888 &  0 &  69 & 188 & $17445^{+2364}_{-2094}$ & $15403^{+1208}_{-1122}$ & $15903^{+1055}_{-991}$  \\
19.5 -- 20.0 & 0.717 &  0 &  78 & 235 & $24423^{+3091}_{-2758}$ & $23846^{+1659}_{-1554}$ & $23987^{+1434}_{-1355}$ \\
20.0 -- 20.5 & 0.439 &  0 &  56 & 173 & $28638^{+4363}_{-3814}$ & $28672^{+2350}_{-2178}$ & $28664^{+2022}_{-1893}$ \\
20.5 -- 21.0 & 0.127 &  1 &  32 & 114 &$56568^{+11880}_{-9952}$ & $65309^{+6709}_{-6107}$ & $63170^{+5672}_{-5223}$ \\
21.0 -- 21.5 & ---   &  0 &   3 &  26 &   --- &   --- &   --- \\
21.5 -- 22.0 & ---   &  0 &   1 &   2 &   --- &   --- &   --- \\
  \noalign{\smallskip}
  \hline
  \noalign{\smallskip}
 \end{tabular}
\end{table*}

\begin{figure*}
\centerline{\psfig{file=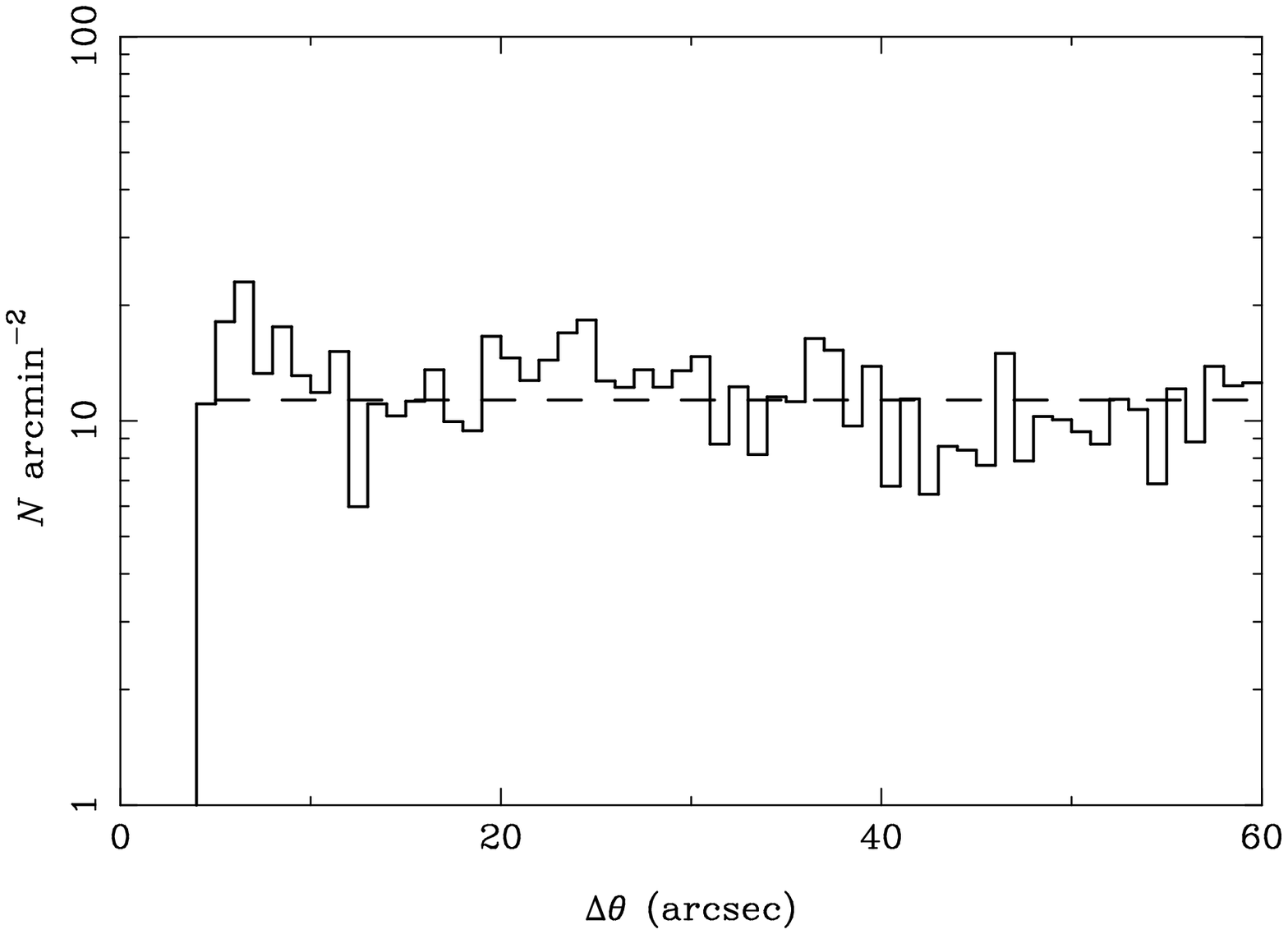,width=8.4cm}
\psfig{file=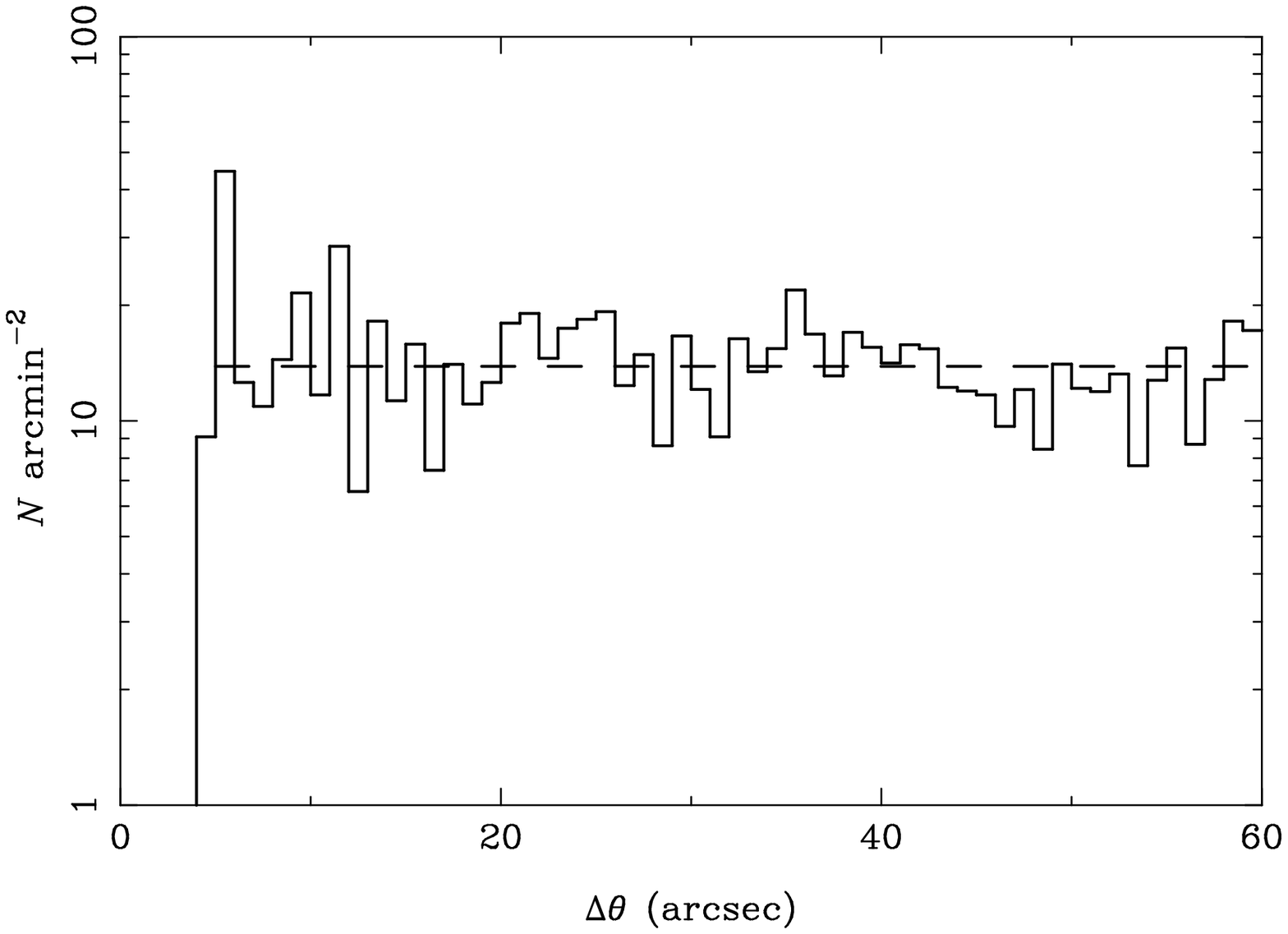,width=8.4cm}}
\caption{Galaxy surface densities as a function of separation from the bright 
star in our non-NVSS fields.  Left panel shows mean for 23 fields within EIS 
Wide patches; right panel shows mean for 14 fields outside of EIS Wide 
patches.  Only for $\Delta \theta < 5\arcsec$ does the star have an effect 
on the detectability of other objects.  Dashed lines indicate the average 
number densities for $5\arcsec \leq \Delta \theta \leq 60\arcsec$.  The slight 
increase in number density at $\Delta \theta < 30\arcsec$ for the EIS 
fields results from our having preferentially imaged fields with one or more 
detected $I$-band sources at $\Delta \theta < 30\arcsec$.}
\label{f-rad}
\end{figure*}

We have used a single set of simulations to determine simultaneously the 
photometric accuracy and completeness limit as a function of magnitude for our 
full source catalogue.  For each field, we extracted a $10\arcsec \times 
10\arcsec$ cutout around every object with $K_\mathrm{s} < 18$ and 
rescaled it by a constant factor so that the source magnitude was shifted into 
the range $16 \leq K_\mathrm{s} < 21$.  Brighter objects with true 
$K_\mathrm{s} < 15$ (about half of which are the central stars themselves) 
were dimmed to magnitudes randomly distributed in the range $16 \leq 
K_\mathrm{s} < 18$; fainter objects with true $15 \leq K_\mathrm{s} < 18$ were 
dimmed to magnitudes randomly distributed in the range $16 \leq K_\mathrm{s} 
< 21$, subject to the constraint of $\geq 1\,\mathrm{mag}$ fading.  We then 
pasted the rescaled cutout back into the original image at a random position 
with separation $5\arcsec \leq \Delta \theta \leq 55\arcsec$ from the central 
star.  We did not consider test separations $\Delta \theta < 5\arcsec$ due to 
the complex coupling of source detection with PSF subtraction at small radii 
(see e.g., Fig. \ref{f-rad}), but avoided separations $55\arcsec < \Delta 
\theta \leq 60\arcsec$ merely to minimize edge effects.  The enforcement of 
$\geq 1\,\mathrm{mag}$ fading ensured that the noise would remain identical to 
that in the original image.  Using the same parameters as in the original 
execution, we ran SExtractor on the modified image and attempted to recover 
the fictitious object within $0.5\arcsec$ of its nominal position.  The 
procedure was repeated 100 times for every $K_\mathrm{s} < 18$ object in each 
of the 42 fields, giving a total of 57,800 iterations.

Figure \ref{f-cpl} and Table \ref{t-counts} show the completeness we derive 
over the range $16 \leq K_\mathrm{s} < 21$ from the dataset as a whole, i.e., 
treating the sum of our 42 $5\arcsec \leq \Delta \theta \leq 60\arcsec$ annuli 
as the ``discrete deep field'' equivalent of a single contiguous $131\,\mathrm{
arcmin}^2$ area with patchy sensitivity.  We estimate our catalogue is 90\% 
complete to $K_\mathrm{s} = 19.2$ and 50\% complete to $K_\mathrm{s} = 20.2$.  
These values include the effects of non-detections due to the proximity of 
another (brighter) object, typically 1--2\% at $K_\mathrm{s} \sim 17$, and 
apply as well to extended as to point sources.  The error bars in Fig. 
\ref{f-cpl} indicate the field-to-field variations in completeness, which 
result from differences in integration time, seeing, and source surface 
density.

Because SExtractor calculates a magnitude for each of the fictitious sources 
it recovers in the 57,800 simulated images described above, we can estimate 
the uncertainty in our photometry from the distribution of differences between 
the recovered sources' nominal and measured magnitudes ($\Delta m \equiv 
m_\mathrm{meas} - m_\mathrm{nom}$).  Empirically, this distribution is 
symmetric about zero (i.e., $<\Delta m> = 0$) down to $K_\mathrm{s} = 20$ with 
large non-Gaussian wings.  At fainter levels, $<\Delta m>$ begins to skew 
negative as sources with $m_\mathrm{meas} > m_\mathrm{nom}$ become 
increasingly difficult for SExtractor to recover.  Figure \ref{f-mag_sdv} 
plots the variance $<\Delta m^2> - <\Delta m>^2$ (after rejection of a few 
catastrophic outliers) as a function of $m_\mathrm{nom}$.  The implied errors 
are significantly larger than the formal uncertainties returned by SExtractor, 
and provide a more realistic indication of the true accuracy of our photometry.

\subsection{Galaxy counts vs. $K_\mathrm{s}$ magnitude}

In order to evaluate the reliability of our source catalogue, we have derived 
a magnitude-counts relation from our data suitable for comparison with the 
published literature.  We begin by using the SExtractor stellarity 
index to perform star/galaxy separation within our source catalogue.  In 
addition to the 240 objects with stellarity index $> 0.97$ that are 
clearly stars, we exclude an additional 151 sources with stellarity index 
above a more conservative threshold of 0.90.  This leaves 1589 sources in the 
42 catalogue fields, of which 3 lie at separations $\Delta \theta < 
5\arcsec$, 448 at $5\arcsec \leq \Delta \theta < 30\arcsec$, and 1138 at 
$30\arcsec \leq \Delta \theta \leq 60\arcsec$.  Division into these three 
subsets isolates the ranges of source separation in which the simulations of 
Sect. \ref{ss-comp} do not constrain the completeness (i.e., $\Delta \theta 
< 5\arcsec$), and in which particularly special care will be needed to 
model AO PSF variation as a function of radius (i.e., $30\arcsec \leq \Delta 
\theta \leq 60\arcsec$; see Steinbring et al. \cite{stei02}).  Table 
\ref{t-counts} bins the data by magnitude, and for the range over which  
we have been able to make reliable completeness corrections ($12.5 \leq 
K_\mathrm{s} < 21$) lists the differential number counts from 
both of the outer annuli and the full $5\arcsec \leq \Delta \theta \leq 
60\arcsec$ range, normalized to units of counts per magnitude per square 
degree.  The $1\sigma$ error bars on the normalized counts are derived from 
Poissonian statistics according to the prescriptions of Gehrels 
(\cite{gehr86}).  

\begin{figure}
\centerline{\psfig{file=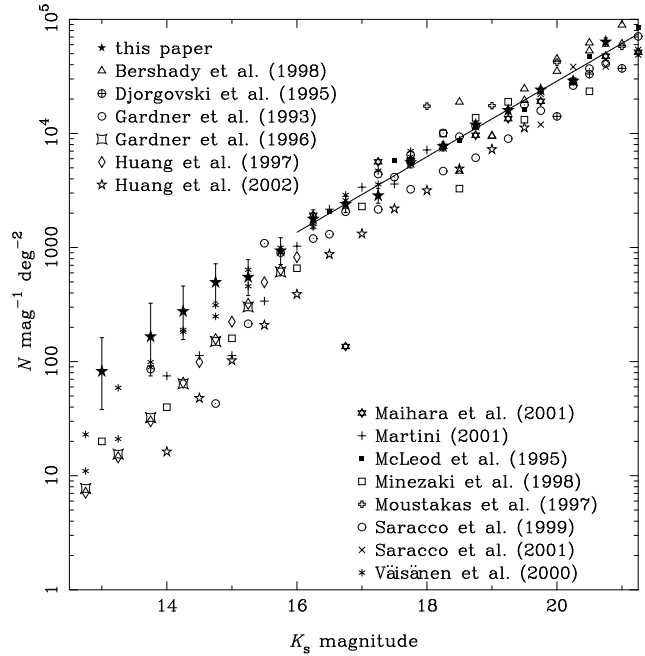,width=8.4cm}}
\caption{Differential galaxy number counts as a function of $K_\mathrm{s}$ 
magnitude, compared to results from the literature.  The solid line indicates 
the best power-law fit to our data (stars with error bars) for $16 \leq 
K_\mathrm{s} < 21$ and $5\arcsec \leq \Delta \theta \leq 60\arcsec$ and has 
a slope $\alpha = 0.33 \pm 0.01$.  References for other symbols are as 
indicated in the figure.}
\label{f-lit}
\end{figure}

Figure \ref{f-mags} plots the completeness-corrected counts for the two 
annuli (as discrete points with error bars) and for the full sample (as a 
continuous curve).  It is immediately apparent-- as from comparison of 
columns (6) and (7) in Table \ref{t-counts}-- that our 42 fields in aggregate 
have a higher surface density of background sources at separations $5\arcsec 
\leq \Delta \theta < 30\arcsec$ from the star than at separations 
$30\arcsec \leq \Delta \theta \leq 60\arcsec$ in the brighter magnitude bins.  
This result is not surprising, given that a number of our fields were selected 
in part precisely because they {\it did} reveal background sources within 
$30\arcsec$ of the central stars in $I$-band images.  In particular, when we 
compare plots of mean galaxy surface density versus angular separation for 23 
fields in EIS Wide patches (most subject to this bias) and for 14 fields 
not in EIS Wide patches or near NVSS sources, as in the two panels of 
Fig. \ref{f-rad}, we see a slight step in surface density at $\sim 30\arcsec$ 
in the former plot that is absent from the latter.\footnote{The slight 
difference in mean surface densities between EIS ($11.3\,\mathrm{arcmin}^{-2}$)
and non-EIS ($13.9\,\mathrm{arcmin}^{-2}$) fields may be due to residual 
contamination by faint stars in the latter, which tend to lie at somewhat 
lower Galactic latitudes.}  Note that this effect arises entirely in the 
bright magnitude bins; at fainter magnitudes ($K_\mathrm{s} \geq 19$), Table 
\ref{t-counts} and Fig. \ref{f-mags} indicate that there is no gradient in the 
source density as a function of source separation.  At the still fainter 
magnitudes that can only be reached with AO-assisted imaging of these fields, 
there should again be no dependence of source surface density on separation 
from the bright star.

Figure \ref{f-lit} shows a comparison of the completeness-corrected galaxy 
number counts calculated from the full $5\arcsec \leq \Delta \theta \leq 
60\arcsec$ sample with comparable results from the literature.  This figure 
includes a line showing the best power-law fit to the number counts for 
the range $16 \leq K_\mathrm{s} < 21$; we derive a slope $\alpha \equiv 
d\,\mathrm{log}\,N/dm = 0.33 \pm 0.01$.  Leaving out the $K_\mathrm{s} \sim 
20.75$ point, which has a large and uncertain correction for incompleteness, 
gives $\alpha = 0.31 \pm 0.01$; recomputing the counts from the 37 
non-NVSS fields only (to circumvent the plausible clustering of the NVSS 
fields' faint sources) gives $\alpha = 0.34 \pm 0.01$.  In all 
cases, the agreement with previously published values in the range $0.32 - 
0.37$ is excellent (e.g., Djorgovski et al. \cite{djor95}; McLeod et al. 
\cite{mcle95}; Bershady et al. \cite{bers98}; McCracken et al. \cite{mccr00}; 
Huang et al. \cite{huan01}).  Our number counts are also in good agreement 
with prior results in terms of normalization for $K_\mathrm{s} \geq 16$ 
(although brighter magnitude bins are affected by the selection bias discussed 
above).  The overall agreement with the literature gives us confidence that 
our approach to source identification is sensible, and (for faint magnitudes) 
that our fields define a representative slice of the extragalactic 
$2.2\,\mathrm{\mu m}$ sky.

\section{Conclusions}

The agreement of the magnitude-counts relation derived from analysis of our 
``discrete deep field'' with previous derivations by other authors is not a 
great surprise.  Indeed, we would expect {\it a priori} that our results 
should be no more biased by cosmic variance than those of groups working with 
contiguous fields of equivalent total area, although the latter will be more 
useful for direct measurements of clustering strengths.  The consistency 
demonstrates that our somewhat {\it ad hoc} approach to field selection-- with 
some dependence on the availability of EIS data and/or the proximity of a 
bright radio or optical source-- has not yielded a biased picture of the deep 
$2.2\,\mathrm{\mu m}$ sky.  As a result, it is possible to proceed to AO 
observations of this set of fields with full confidence that any conclusions 
based on the aggregate properties of their faint near-IR sources will be 
statistically sound.  We hope that the southern hemisphere user community will 
find the catalogue of 391 faint stars and 1589 galaxies we have provided here 
to be useful in fully exploiting the use of 8m-class telescopes at or near 
their diffraction limits.

\begin{acknowledgements}
We thank the staff at the NTT, in particular support astronomers St{\' e}phane 
Brillant and Pierre Leisy, for their advice on strategy and help in conducting 
the observations.  We are grateful to Jo{\~ a}o Alves and Charlie Lada for 
executing the 7 March 2001 observations on our behalf as part of a time swap, 
and to H{\' e}l{\` e}ne Dickel and James Larkin for helpful suggestions.  
M.~J.~J. has been supported by the European Research and Training Network on 
the Physics of the Intergalactic Medium.  This research has made use of NASA's 
Astrophysics Data System Bibliographic Services, and of the NASA/IPAC 
Extragalactic Database (NED), which is operated by the Jet Propulsion 
Laboratory, California Institute of Technology, under contract with NASA.

\end{acknowledgements}

\end{document}